 \documentclass[final,3p,times]{elsarticle}

\usepackage{amssymb}
\usepackage{amsmath}
\usepackage{lineno}
\usepackage{booktabs}
\usepackage{mathtools}
\usepackage[section]{placeins}
\usepackage{natbib}
\usepackage{textcomp}

\journal{}

\begin{document}

\begin{frontmatter}



\title{Inter-Modal Raman Amplification in Space-division Multiplexed Systems} 


\author[aff1]{Mario Zitelli}
\author[aff2]{Louis Andreoli}
\author[aff2]{Claire Autebert}
\author[aff2]{Jean-Philippe Gauthier}
\author[aff2]{Guillaume Labroille}
\author[aff1]{Stefan Wabnitz} 

\affiliation[aff1]{organization={Università di Roma Sapienza, Dept. DIET},
            addressline={Via Eudossiana 18}, 
            city={Rome},
            postcode={00151}, 
            country={Italy}}
\affiliation[aff2]{organization={CAILabs},
addressline={38 boulevard Albert 1er}, 
city={Rennes},
postcode={35200}, 
country={France}}

\begin{abstract}
We theoretically analyze and experimentally demonstrate the possibility of amplifying optical signals in an unrepeatered mode-division multiplexed tranmsmission system, through inter-modal stimulated Raman scattering process between signal and pump beams coupled onto distinct modes of a few-mode graded-index optical fiber. 
\end{abstract}





\end{frontmatter}



\section{Introduction}
\label{sec1}

In recent years there is much research interest in methods for solving the so-called capacity crunch of long-distance optical fiber transmission systems. Unrepeatered cables using single-mode fibers are capable of an overall capacity (bit/s) of $C_{tot}=N_\lambda N_f C$, with $N_f$ the number of fibers in the cable, and $N_\lambda$ the number of wavelength-division multiplexed (WDM) channels in each fiber; $C=B \cdot log _2 \left(1+SNR\right)$ is the maximum Shannon capacity of a fiber channel, with $B$ the optical channel bandwidth (Hz) and $SNR$ the signal-to-noise ratio at the receiver. A further capacity increase needs to address new forms of channel multiplexing and optical propagation, such as space-division multiplexing (SDM), whereby different optical channels are multiplexed over the different cores of a multicore fiber (MCF), or modes of a multimode fiber (MMF) \cite{Essiambre_5420239}. The potential of MMFs for increasing the transmission capacity of long-distance optical links, and the increased tolerance to high signal powers with respect to single mode fibers, have attracted considerable interest in recent years \cite{Richardson-NatPhot-2013-94-2013,Agrell_2016}.

In this context, a major issue is the development of optical amplifiers which are compatible with SDM. So far, most attention has been dedicated to developing multimode rare-earth doped fiber amplifiers. In contrast, we would like to explore the possibility of exploiting MMF amplifiers based on Raman gain for SDM systems. Raman amplification in multimode waveguides was proposed by several patents and papers. In the patent \cite{Rice_2000}, a Raman fiber amplifier was proposed, involving the injection of a signal beam in a multimode inner core of a dual-clad optical fiber. Whereas the pump power is coupled to the multimode outer core, with an outer cladding confining the pump power in the outer core. In this amplifier, signal and pump are injected into different cores, with no modal multiplexing, nor modal selection at the input. In Ref. \cite{Polley_4077082}, Raman amplification of a signal in graded-index (GRIN) MMF was investigated, confirming that the radial dependence of the Raman gain coefficient promotes the amplification of the fiber’s transverse fundamental mode $LP_{01}$ with respect to the higher-order modes (HOM). In that work, a signal at 10 Gbit/s repetition rate at 1550 nm wavelength, and a continuous wave (CW) pump at 1455 nm, were both injected into the MMF by means of a standard single-mode fiber (SMF). However, no mode-division multiplexing nor a selective use of the fiber modes for signals and pumps were used. In a conference paper \cite{Krummrich:11}, multimode and multicore fiber pumping schemes were proposed for Raman amplification, based on modal separation and coupling using fiber tapers and GRIN lenses. The pump radiation was injected by using dichroic mirrors in free-space propagation, after modal separation by fiber tapers and GRIN lenses. However, the pump radiation was not directly injected on selected fiber modes by using mode-division multiplexers.

In this work, we numerically and experimentally investigate the possibility of amplifying signals in a GRIN few-mode fiber (FMF), by multiplexing both signal and pump beams on different fiber modes. Signal amplification at the expense of the pumps is produced via inter-modal stimulated Raman scattering (IM-SRS), which is distributed along the fibre. The injection of signals and pumps, and the extraction of the signal modes, could be performed by mode-division multiplexers based on the multi-plane light conversion (MPLC) technology \cite{LABROILLE201793}. 
Thanks to this architecture, multiple processes of Raman amplification act simultaneously, which increases the overall amplification efficiency. 

In addition, MMF may carry several tens of Watt of optical pump and signal power; the use of multiple IM-SRS processes combined with high pump power levels could enable the extension of unrepeatered fiber links well above the limit of single-mode systems, as well as the deployment of long-haul systems with hundreds of fibers into a single submarine/terrestrial cable.

\section{Numerical Simulations}

We carried out extensive numerical simulations of signal amplification based on IM-SRS in MMF, by using an advanced numerical model based on coupled-mode generalized nonlinear Schrödinger equations (GNLSE) \cite{Poletti:08, Wright2018a}, modified to include the presence of random modal coupling (RMC). 

\begin{multline}
\frac{\partial A_p\left(z,t\right)}{\partial z}=i\left(\beta_0^{\left(p\right)}-\beta_0\right)A_p-\left(\beta_1^{\left(p\right)}-\beta_1\right)\frac{\partial A_p}{\partial t}+i\sum_{n=2}^{4}{\frac{\beta_n^{\left(p\right)}}{n!}\left(i\frac{\partial}{\partial t}\right)^nA_p} -\frac{\alpha_p\left(\lambda\right)}{2}A_p
\\ +i\sum_{m}{q_{mp}A_m}+i\frac{n_2\omega_0}{c} \sum_{l,m,n} Q_{plmn} \{\left(1-f_R\right)A_lA_mA_n^\ast +f_RA_l\left[h\ast\left(A_mA_n^\ast\right)\right]\}    .
\label{eq:GNLSE}
\end{multline}

\noindent The right-hand side terms of Eq. \ref{eq:GNLSE} describe: modal dispersion, four orders of chromatic dispersion, wavelength-dependent losses, RMC with coefficients $q_{mp}$, nonlinear Kerr and Raman terms, respectively. 
The $Q_{plmn}$ are cross-terms corresponding to the inverse of the effective modal areas, providing appropriate weights to inter-modal four-wave mixing (IM-FWM) and IM-SRS terms. 
The Raman term in Eq. \ref{eq:GNLSE} is responsible for signal amplification, and it contributes with a fraction $f_R=0.18$; the expression $h\ast\left(A_mA_n^\ast\right)$ denotes time convolution with the Raman response function $h(t)$, with typical time constants of 12.2 and 32 fs \cite{Stolen1989}. The coefficients $q_{mp}\left(z\right)$ account for linear RMC, and are associated with random imperfections of the refractive index profile due to microbending \cite{Ho:14}.  Here we suppose that RMC only occurs among degenerate modes.

The parameters of the FMF at the wavelength $\lambda=1460$ nm ($1550$ nm) are: $\beta_2=-18.0 (-27.0)$ ps$^2$/km, $\beta_3=0.11 (0.14)$ ps$^3$/km, $\alpha=1.14\times10^{-4}$ ($5.0\times10^{-5}$) m$^-1$, $n_2=2.6\times10^{-20}$ m$^2$/W, fiber radius $a=14$ $\mu$m. 
The injected signals are composed by return-to-zero (RZ) encoded pulse sequences with 2 ps pulsewidth and repetition rate of 40 Gb/s, launched over one or more fiber modes at 1550 nm wavelength; the total signal power is 20 mW, corresponding to 1 pJ of energy per pulse. The pumps are CW beams at 1460 nm, distributed over a variable number of modes (5 to 14).

\begin{figure}[h]
\includegraphics[width=0.7\textwidth]{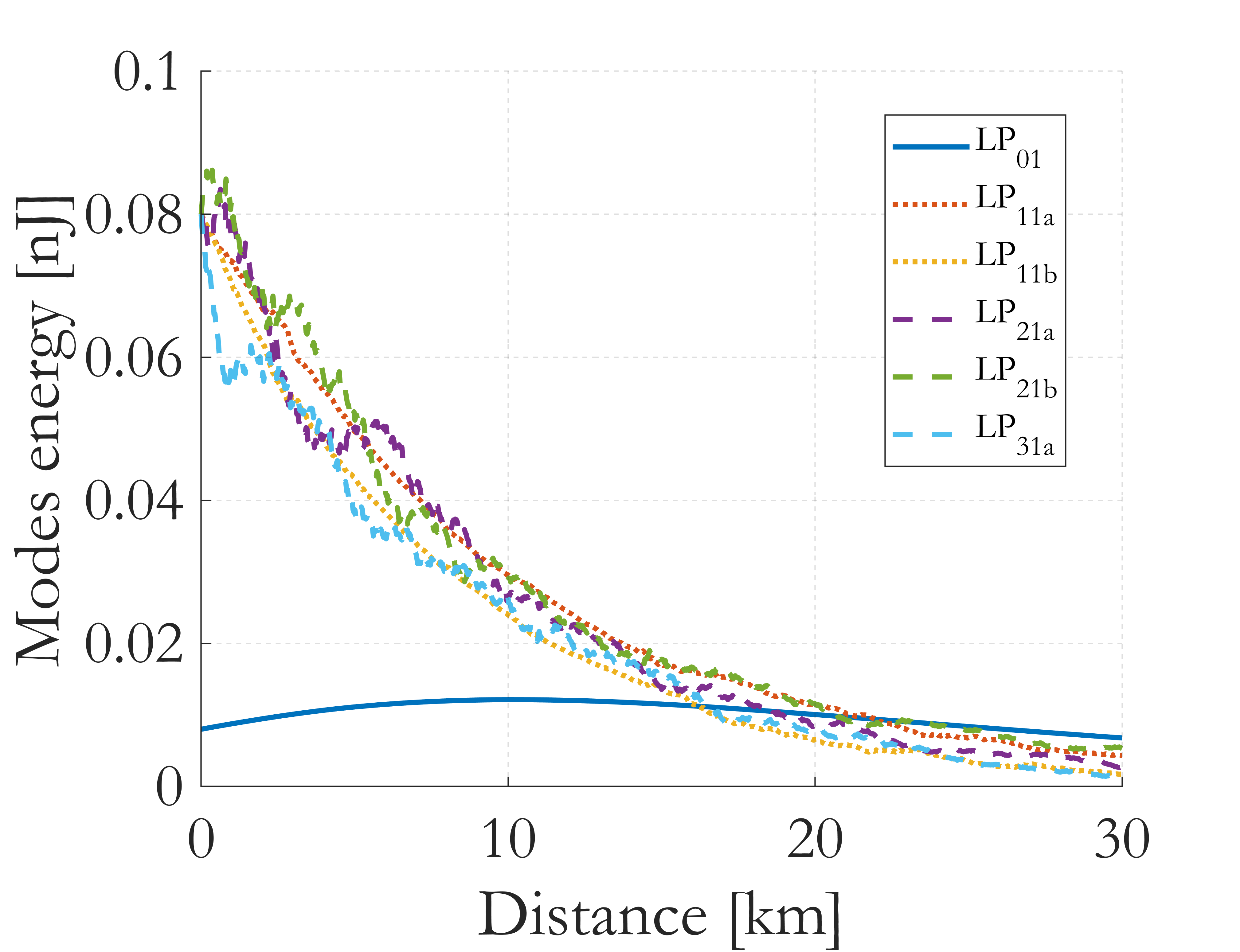}	\centering	
\caption{Numerical simulation with one signal and 5 pump modes (1 W total pump power).}
\label{fig:Fig1}
\end{figure}

\begin{figure}[h]
\includegraphics[width=0.7\textwidth]{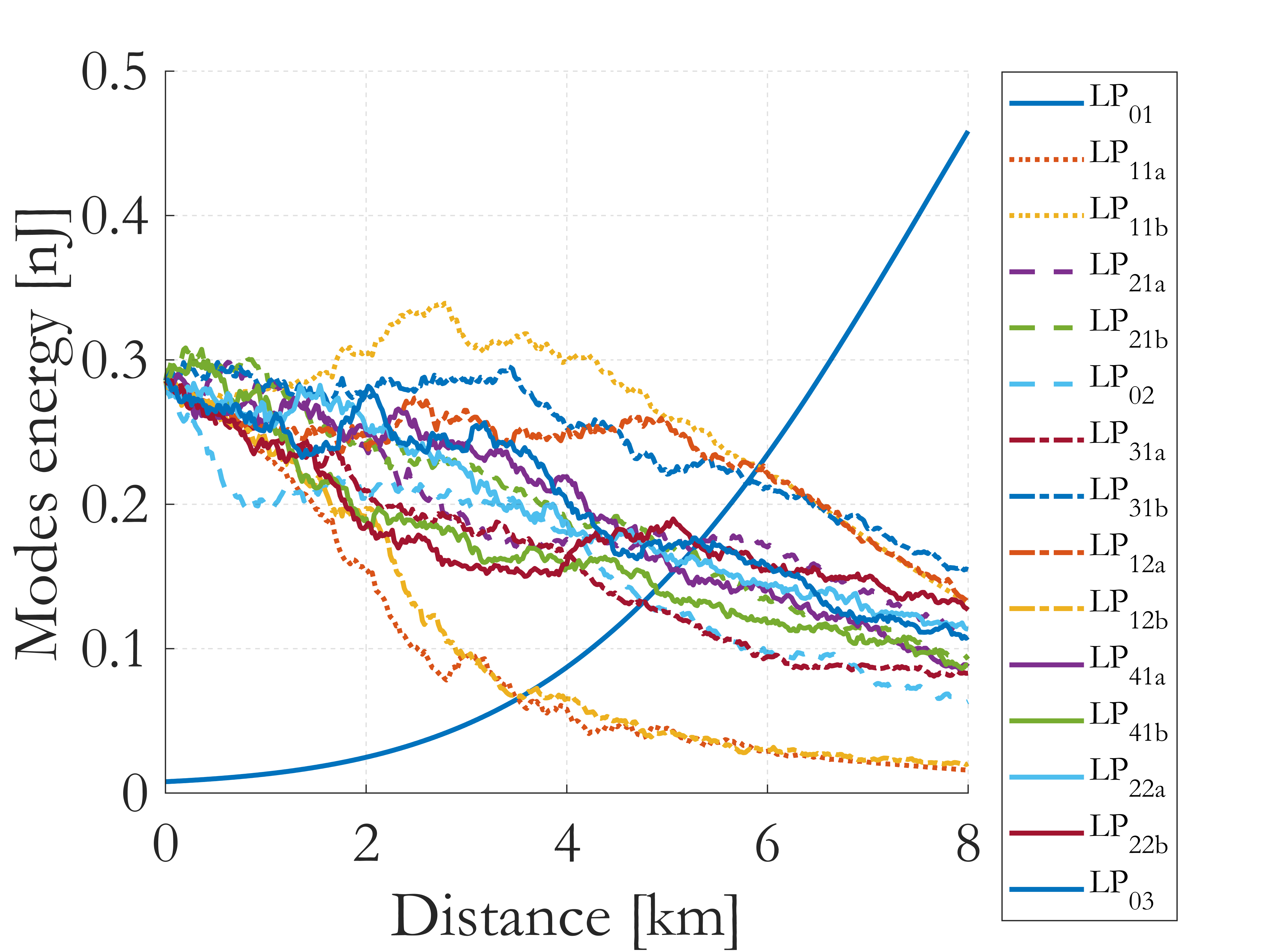}	\centering	
\caption{Numerical simulation with one signal and 14 pump modes (10 W total pump power).}
\label{fig:Fig2}
\end{figure}

Figure \ref{fig:Fig1} illustrates the numerically computed evolution of the signal and pumps energy, calculated over the simulation window (400 ps), vs. transmission distance. Six pump beams were launched in modes 2-5 ($LP_{11a}$ to $LP_{31a}$), for a total power of 1.0 W equi-distributed over these modes. In terms of energy, 8 pJ for the signal, and 67 pJ for each pump is transmitted in the simulation window.
RMC is responsible for the observed random energy fluctuations of the pumps, while the pump energies globally decrease because of power transfer to the signal, and linear losses. The signal on-off gain induced by IM-SRS, defined as the ratio of the output signal power with and without the pumps, overcomes the fiber losses when the total injected pump power is 1 W. The obtained gain is $G=3.4$ dB after 8 km, and reaches for a maximum of $G=3.9$ dB at 10.1 km; after this distance, pump depletion and absorption prevents further amplification. 
When the pump power is increased up to 10 W, and it is distributed over 14 fiber modes, the signal gain can easily reach up to 19.3 dB at 8 km of distance, as it is illustrated in Fig. \ref{fig:Fig2}. A similar gain is obtained when injecting multiple signals, multiplexed over the modal channels which are not used for the pumps.

\section{Experimental Results}

\begin{figure}[h]
\includegraphics[width=1\textwidth]{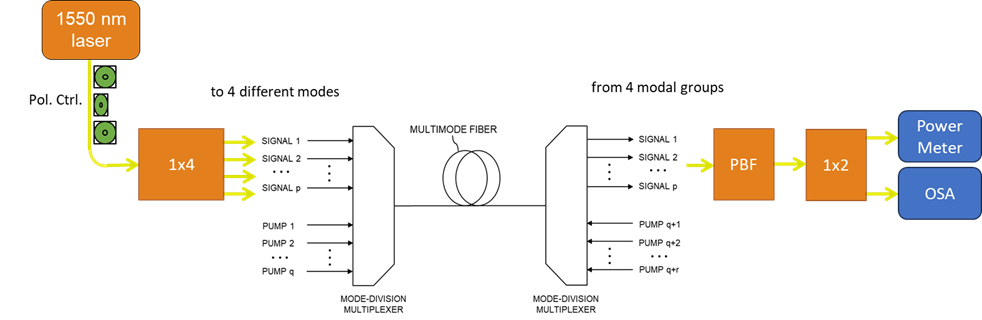}	\centering	
\caption{Experimental setup.}
\label{fig:Fig3}
\end{figure}

The experimental setup for demonstrating distributed Raman amplification using mode-multiplexed signals and pumps is illustrated in Fig. \ref{fig:Fig3}. Pumps are injected at the transmitter or at the receiver sides, following the design proposed in \cite{Zitelli_pat2023} and optimized according to the numerical simulations presented in the previous section. In the proposed architecture, no WDM couplers are used to inject the signals and the pumps into the FMF. On the other hand, signals and pump are directly applied to the modal multiplexers, and coupled to distinct modes.

The device used for multiplexing the fiber and pump modes is a custom-designed (in order to minimize insertion losses for both Raman pumps and signals in the C band) Cailabs unit (PROTEUS-C-15-1550-CUSTOM), supporting the Hermite-Gauss modes $HG_{00}$, $HG_{01}$, $HG_{10}$, $HG_{02}$, $HG_{11}$, $HG_{20}$, $HG_{03}$, $HG_{12}$, $HG_{21}$, $HG_{30}$, $HG_{04}$, $HG_{13}$, $HG_{22}$, $HG_{31}$, $HG_{40}$ of a Prysmian GRIN-LP9 FMF.
Four signals at 1550.1 nm were polarization controlled, split into 4 beams and coupled to modes $HG_{00}$ (modal group 1), $HG_{11}$ (group 3), $HG_{21}$ (group 4), $HG_{31}$ (group 5) of the 8 km long Prysmian GRIN-LP9 FMF; the signal modes are non-degenerate, and belong to different modal groups. Five pump lasers with wavelengths ranging between 1460 and 1470 nm were injected into modes $HG_{01}$, $HG_{02}$, $HG_{03}$, $HG_{13}$, $HG_{22}$ at the input (codirectional pumping) or at the output (contradirectional pumping) side of the system, respectively.

At the 1550 nm wavelength, the system (that is, including the MUX pair and the transmission fiber) average (worst) modal cross-talk among different modal groups was measured to be equal to -15.3 dB (-10.7 dB), and the system insertion loss was 10.4 dB (11.8 dB).
At 1460 nm, the average (worst) cross-talk was -17.8 dB (-14 dB), and the insertion loss 11.1 dB (12.2 dB). The modal insertion loss ascribed to a single multiplexer was 3.2 dB at 1460 nm, and 3.3 dB at 1550 nm.

At the output of the modal demultiplexer, the degenerate modes of a same group were added by using a $N\mathrm{x}1$ coupler (mode-group demultiplexing), being $N$ the number of degenerate modes of the measured modal group. Hence, signal modal channels were added to the respective degenerate modes, in order to eliminate the effects of intra-group modal cross-talk, and passed through a 1 nm passband optical filter (PBF) in order to cut off the pump power. A $1\mathrm{x}2$ splitter provided the individual modal groups to an optical powermeter with $\mu$W resolution and an optical spectrum analyzer (OSA) with 0.02 nm resolution (Yokogawa AQ6370D).

In a first test, co-propagating signals and pumps were used. Five pump lasers with wavelengths ranging between 1460 and 1470 nm were injected in modes $HG_{01}$ (group 2), $HG_{02}$ (group 3), $HG_{03}$ (group 4), $HG_{13}$ and $HG_{22}$ (group 5) at the input side of the system. The pump power was equi-distributed among modes, and it was adjusted in order to couple 1 W of total power in the FMF. Table \ref{tab1} reports the obtained gains for the four signals in the co-propagating configuration; signals were simultaneously propagating in the MMF. 
The obtained gains, with an average of 2.6 dB, are in relatively good quantitative agreement with their numerical predictions, and fully overcome the fibre losses.

\begin{table}[h]
\caption{}
\label{tab1}  
\centering	
\begin{tabular}{@{}lllllll@{}}
\hline
Input & Output & Signal & Pump modes & Pump & Signal Gain & Signal Gain\\
signal & signal & power &   & power & Co-propagating & Counter-propag.\\
modes & groups & [mW] &   & [W] & [dB] & [dB]\\
\hline
$HG_{00}$ & 1 & 2.5 & $HG_{01}$, $HG_{02}$, $HG_{03}$, $HG_{22}$, $HG_{13}$ &	1 &	2.45 & 1.33\\
$HG_{11}$ & 3 & 2.5 & $HG_{01}$, $HG_{02}$, $HG_{03}$, $HG_{22}$, $HG_{13}$ &	1 &	2.35 & 2.14\\
$HG_{21}$ & 4 & 2.5 & $HG_{01}$, $HG_{02}$, $HG_{03}$, $HG_{22}$, $HG_{13}$ &	1 & 2.51 & 2.16\\
$HG_{31}$ & 5 & 2.5 & $HG_{01}$, $HG_{02}$, $HG_{03}$, $HG_{22}$, $HG_{13}$ &	1 & 2.90 & 1.95\\
\hline
\end{tabular}
\end{table}

\begin{figure}[h]
\includegraphics[width=0.7\textwidth]{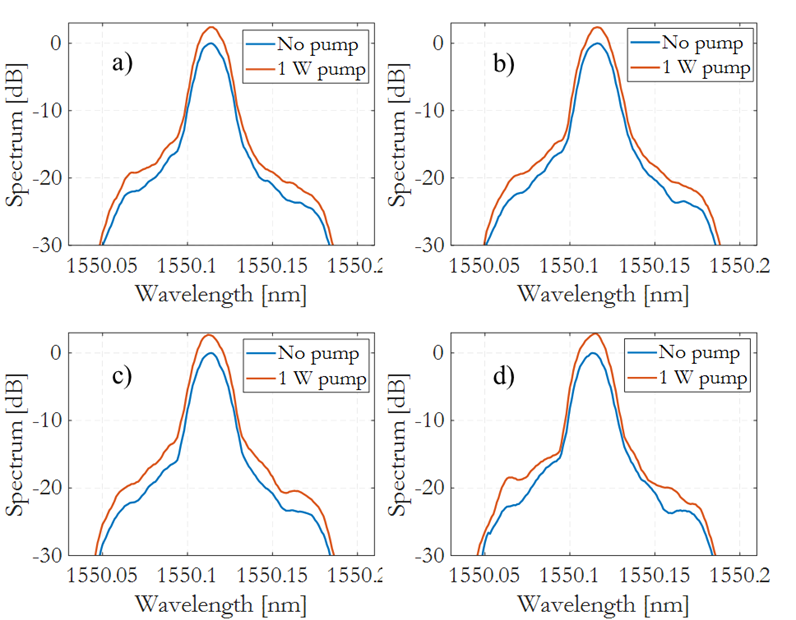}	\centering	
\caption{Measured spectra of the 4 signals, with and without 1 W of pump power applied to 5 modes.}
\label{fig:Fig4}
\end{figure}

Figs. \ref{fig:Fig4}(a, b, c, d) show the spectra of the signals from modal groups 1, 3, 4, 5, respectively, with and without the injection of pump power. The optical noise is not visible in the figures; however, it was measured an optical signal-to-noise ratio $OSNR$ ranging between 50 dB and 52 dB for the different output modal groups.

In a second test, counter-propagating signals and pumps were used. Five pump lasers with wavelengths ranging between 1460 and 1470 nm were coupled to modes $HG_{01}$ (group 2), $HG_{02}$ (group 3), $HG_{03}$ (group 4), $HG_{13}$ and $HG_{22}$ (group 5) from the output side of the system. Hence, signals and pumps propagate in opposite directions. Output signals were extracted by using a $N\mathrm{x}1$ coupler adding the degenerate modes of a group; modes used for pump injection were not added to the output coupler. Hence, the output signals were composed by: 
group 1: $HG_{00}$;
group 3: $HG_{11}+HG_{20}$ (no $HG_{02}$);
group 4: $HG_{12}+HG_{21}+HG_{30}$ (no  $HG_{03}$);
group 5: $HG_{04}+HG_{13}+HG_{22}+HG_{40}$ (no $HG_{22}$, $HG_{13}$).
Also in this case, signals were passed through a band-pass filter which eliminated the pump residual, before measuring power and spectrum.

Table \ref{tab1} reports the obtained signal gains in the counter-propagating configuration; also in this case, signals were simultaneously propagated over different modes. The measured average gain is now 1.89 dB, which is 0.7 dB lower than in the co-propagating case. We did not test the counter-propagating case by numerical simulations, since our model only allows for studying the forward propagation of both pumps and signals.

\section{Discussion and Conclusions}

The capability of obtaining Raman gain for SDM signals was numerically and experimentally demonstrated, when signal beams at a wavelength of 1550.1 nm and pump beams at a wavelength between 1450 and 1470 nm are coupled into different modes of a GRIN FMF, by means of a modal multiplexer based on the MPLC technology. Signal amplification in the transmission fiber, at the expense of the pumps, was obtained via IM-SRS, which is distributed along the fiber. At the receiver, the signal modes were retrieved via a corresponding modal demultiplexer. The measured gain entirely compensated for the fibre losses experienced by the mode-multiplexed signals.
A larger gain (2.6 dB on average) was observed when using the co-propagating architecture of signal and pump modes, when compared with the counter-propagating case (1.9 dB of signal gain in average).

Numerical simulations confirmed the experimental data, and suggested that, by increasing the pump power to 10 W distributed over a larger number of modes, the signal gain could increase up to 19.3 dB over 8 km of GRIN FMF. The injection of such pump power is feasible as far as the MMF is concerned, but it poses challenges to the mux/demux design.

The successful demonstration of the IM-SRS architecture could enable the use of high-capacity, optical unrepeatered systems over long-haul systems, which are typical of international transport networks. In turn, this may permit to install long-haul cables including hundreds of MMFs, carrying no electrical current for unwanted in-line repeaters. 

\vspace{5mm} 
Funding: 

HORIZON EUROPE European Research Council Proof-of-Concept project MULTIBRIDGE (101081871).

The work of S.W. was also supported by the EU under the NRRP of NextGenerationEU, partnership on “Telecommunications of the Future” (PE00000001 - “RESTART”).











\end{document}